# Suspended Graphene Ballistic Josephson Weak Links

Naomi Mizuno, Bent Nielsen, and Xu Du[*]

Department of Physics and Astronomy, Stony Brook University, Stony Brook, NY 11794-3800, USA

## Abstract

The interplay of graphene and superconductivity has attracted great interest for understanding the two-dimensional Dirac Fermion physics and for superconducting device applications. In previous work, graphene-superconductor junctions fabricated on standard $SiO_2$ substrates were highly disordered. Due to charge carrier scattering and potential fluctuations, the expected intrinsic properties of graphene combined with superconductors remained elusive to experimental observations. We propose suspended graphene-superconductor junctions as an effective solution to the problem. Here, we show the fabrication and characterization of the high-quality suspended monolayer graphene-NbN Josephson junctions that have device mobility greater than 150,000 $cm^2/Vs$, with minimum carrier density below $10^{10}\,cm^{-2}$ and a formation of supercurrent at temperature above 2 K. The devices have exhibited a ballistic Josephson current that depend linearly on the Fermi energy of graphene, a manifestation of the "ultra-relativistic" band structure of Dirac electrons. This work paves the way for future studies of the intrinsic Dirac Fermion superconducting proximity effect and new graphene-superconductor hybrid junctions.

[*]Correspondence to: xu.du@stonybrook.edu

Graphene is a promising material for building superconducting devices. In ambient environment, it is inert against surface oxidation which makes it relatively easy to fabricate transparent interface with the superconducting electrodes. The low charge carrier density inside graphene is easily tunable using electric field effect, adding a useful knob for studying the density/energy dependence of the physical properties of the system. Moreover, the quasiparticles in graphene are Dirac fermions with linear energy dispersion and a vanishing density of state (DOS) at the Dirac point where the conduction and valence bands touch. Such electronic structure offers intriguing new physics in superconducting proximity effect[1-5]. In particular, at close vicinity to the Dirac point, the specular Andreev reflection[1] and a supercurrent supported[2] by evanescence modes have been theoretically predicted. Away from the Dirac point, the intrinsic ballistic Josephson current depends linearly on the Fermi energy[2]. From the practical point of view, the high mobility of the charge carriers in graphene allows a decent magnitude of supercurrent in the superconducting weak links. The tunability of the supercurrent through electric field gating suggests the potential applications in the graphene-superconductor devices[6-9].

Graphene Josephson device was first demonstrated in the graphene-Al junctions[3,4] that were characterized at millikelvin temperatures. More recently, graphene Josephson devices have been studied using superconducting contacts with higher transition temperatures, such as Pb[10], Nb[11,12] and ReW[12]. Up to the present, the reported graphene-superconductor devices have been fabricated on $SiO_2$. A serious limitation of the devices is that the substrates induce severe disorders from the roughness and charge traps. Such disorders limit the devices to be within the diffusive regime with a typical electron mean free path of a few tens of nanometers. Random doping from the substrate creates the electron-hole puddles[13] and limits the minimum carrier density to ~$10^{11}$ cm$^{-2}$, which corresponds to large potential fluctuations of > 30 meV. This hinders the tuning of the Fermi energy of graphene to the close vicinity of the Dirac point.

To date, there have been two major approaches for fabricating high mobility graphene devices: by suspending graphene[14,15] and by placing graphene on a h-BN substrate[16]. Compared to the reported graphene-hBN devices, suspended graphene devices are restricted in the channel size and device geometry, and yet they give the highest device mobility with the lowest potential fluctuations (minimum residue carrier density). The maximum mobility achieved in the suspended graphene junctions is much higher than 200,000 cm$^2$/Vs for a minimum carrier density achievable at ~$10^9$ cm$^{-2}$. For the Josephson junction devices, only simple two-terminal geometry with short channel length is necessary. Therefore, suspended graphene technique is in principle suitable.

Fabrication of the suspended graphene-superconductor junction has two main challenges: The first is to make the suspended structure without wet chemical etching (typically requires hydrofluoric acid or buffered oxide etchant[14,15] which attacks most superconductors), and the second is to produce highly transparent graphene-superconductor contacts. To fabricate the suspended graphene devices, we follow the procedure illustrated in Fig. 1. First, we coat a Si/SiO$_2$ substrate with a polymethyl methacrylate (PMMA) 950K A4 resist spun at 3000 RPM. The thickness of the resist spacer (~250nm) determines the distance between the suspended graphene channel and the surface of SiO$_2$ (see Supplementary Information). Graphene is then mechanically exfoliated on the PMMA film, followed by spin coating of a methyl methacryllate (MMA) EL8.5 copolymer resist for the contact definition (Fig. 1a). By the electron beam dose control, we expose and develop the 3-dimensional surface profile that forms a suspended structure after the metallization and lift-off. A low E-beam dose is used for the exposure of the contact area on top of graphene, and a high dose is used for the surrounding areas of the contacts. Consequently, after developing process, both of the exposed MMA/PMMA layers are removed from the substrate, while only the MMA will be removed from the graphene contact area (Fig. 1b). It should be noted that for the E-beam exposure an accurate dose control is not necessary; because as long as the dose is sufficiently high to expose through the MMA layer, the graphene flake itself will block the developer, preventing the overdeveloping of the PMMA underneath. This gives a high fabrication yield at relatively mild demands on the exposure parameters. Since the graphene-covered plateau is high above the substrate surface, in order to ensure the continuity of the metallic contacts we apply a gradual exposure dose change between the high dose and low dose regions. This simple technique allows the fabrication of suspended graphene junctions with arbitrary metal contacts and on arbitrary substrates. Comparing with the wet etching method with similar Au/Ti contacts[14,15], this method similarly yields high-quality suspended graphene devices with maximum mobility > 200,000 cm$^2$/Vs at minimum carrier density of ~ 3 x 10$^9$ cm$^{-2}$. Well-defined integer quantum Hall plateaus of ν = 2, 6, 10, 14 can readily be observed in a low magnetic field of 0.5 T at 4.2 K (see Supplementary Information).

Following the E-beam lithography process, the sample is metalized in a high vacuum chamber with a base pressure < 2 x 10$^{-8}$ Torr, equipped with a four-pocket e-beam evaporator and a DC magnetron sputtering source. We reactively sputter a superconducting NbN film using Nb target in Ar/N$_2$ plasma. Careful adjustments of the N$_2$ and Ar gas partial pressure, the DC sputtering power and the sample-target distance are carried out to minimize the stress, as well as optimizing the transition temperature of the thin film. In addition, prior to the sputtering of NbN, we e-beam pre-evaporate a buffer layer of Ti (~2 nm)/Pd (~1.5 nm) to create a highly transparent interface between graphene and the NbN contacts and to reduce damage from energetic ions. The Ti layer serves the purpose of improving the adhesion of the contact material with graphene and the Pd layer prevents undesirable reaction of the Ti with the N$_2$ plasma. Immediately after E-beam evaporation of the Ti/Pd bilayer, we sputter NbN without breaking the vacuum (Fig. 1c). To start the reaction, the pressures of Ar (6.7 mTorr) and N$_2$ (0.9 mTorr) are independently controlled by adjusting their corresponding flows. The sample is located 10 cm away from a 3'' Nb target. A constant power of 470 W is maintained during the sputtering. After the plasma is started, we obtain and stabilize sputtering voltage of ~315 V and current of ~1.49 A during 80s pre-sputtering, and then we sputter 70 nm thickness of NbN onto the sample at the deposition rate of ~1nm/s for 80s. The lift-off is performed in acetone after the metallization. The sample is immersed in two successive acetone baths at 80℃. Then, while the sample is kept wet, it is transferred and rinsed in a room temperature IPA Bath. Finally, the sample is quickly moved into a hot IPA bath, which is slightly below the boiling point, and then directly taken out to be dried in air (Fig. 1d).

Combining the wet-etching-free suspending techniques and the NbN deposition, we fabricated the monolayer graphene-NbN weak link (Fig. 1e). Here the graphene weak link has a width of 1.8 μm and a length of 0.5 μm. The device was cooled in a cryostat for electrical characterizations. The as-fabricated suspended graphene device typically does not show good quality before the current annealing treatment (Fig. 2a Inset). A large current (~ 0.3 - 0.4 mA/μm) annealing was conducted to clean the graphene channel *in situ*. The NbN leads became superconducting below the transition temperature of $T_c \sim 12$ K, and we observed a decreasing weak-link resistance with decreasing temperature. To monitor the effect of large current annealing on the graphene-NbN interface, we measure the Andreev reflection features after each high current annealing ramp. Fig. 2b shows the typical differential resistance ($dV/dI$) vs. bias voltage, measured at 4.2 K. The features of the multiple Andreev reflections are clearly observable for the as-fabricated device. We notice that the superconducting gap ($2\Delta \sim 1 meV$) appears to be rather strongly suppressed compared to the estimation from the transition temperature ($\Delta(0) = 1.764 k_B T_c \sim 1.8 meV$). Similar situations had been previously observed in other Josephson weak links where Pd was used between the channel and the superconducting leads[17]. Determination of how to reduce such gap suppression requires further study. At near-zero bias a sharp dip can be seen corresponding to the development of the supercurrent. While current annealing, the device is Joule-heated under the increasingly higher current. The mobility of the device increases while the ratio of the zero-bias resistance to the normal state resistance decreases, indicating the high transparency of the graphene-NbN interface. The dV/dI vs. $V_b$ curve develops into a "V" shape after the annealing, and the oscillatory multiple Andreev reflection feature prior to current annealing becomes less pronounced. The observed changes in the bias voltage dependence of the sub-gap differential resistance from current annealing are qualitatively consistent with a reduction of scattering centers inside the graphene weak link, and possibly at the interface. It has been shown theoretically[18] that for a single quantum channel, the oscillatory features associated with the multiple Andreev reflections are enhanced by a finite reflection probability. For graphene with two dimensional modes, we can expect the modes to transmit current independently with their respective transparency. For near ballistic junctions, with reduction of scattering and hence enhancement of junction transparency, we expect a reduction in the oscillatory amplitude for multiple Andreev reflections. After the current annealing, the maximum mobility of the device shown here reaches $\mu \sim 150{,}000$ cm$^2$/Vs for the electron branch, with a minimum carrier density of $\sim 5 \times 10^9$ cm$^{-2}$ (Fig. 2a). The carrier mobility in this electron branch follows $\mu \sim n^{-1/2} \sim 1/E_F$ (i.e. conductivity $\sigma \sim n^{1/2} \sim E_F$), which is consistent with ballistic transport in a two terminal device[19,20].

We further cool the sample to below 4.2 K. A clear evidence of the Josephson current is observed below 3 K. The IV characteristics of the junction measured at T = 2.2 K is shown in Fig. 3b. By changing the gate voltage we can tune the magnitude of the supercurrent. Fig. 3b shows the temperature dependence of the Josephson current measured at $V_g$ = 5 V. With increasing temperature the supercurrent decreases and a zero-bias resistance becomes evident at T ~3 K. Based on the junction parameters, we estimate (for $V_g$ = 5V, $R_N$ = 500 Ω, $I_c$ ~100 nA, C~ $10^{-12}$ F) the Stewart-McCumber parameter to be

$$\beta_c = \frac{2\pi I_c R_N^2 C}{\Phi_0} \sim 80 \gg 1.$$ (Here the value of C is estimated from the capacitance between the source and drain pads coupled from the conducting back gate[3] ). This indicates the device to be underdamped based on the Resistively and Capacitively Shunted Junction (RCSJ) model[21], in which the Josephson weak link is described by a simple circuit consisting of a "intrinsic" Josephson junction, a capacitor, and a resistor in parallel. In a current driven Josephson weak link, the RCSJ can be described by equation:

$$I = \sin\phi + \frac{\hbar}{2eRI_c}\frac{d\phi}{dt} + \frac{\hbar C}{2eI_c}\frac{d^2\phi}{dt^2},$$ where $i$ is the driving current, $\phi$ the phase difference between the two superconducting contacts, $R$ the normal state resistance, $C$ the junction capacitance, and $I_c$ the intrinsic Josephson current. For a underdamped Josephson weak link, RCSJ model yields IV characteristics which are highly hysteretic. On the other hand, the IV curves at the temperatures and gate voltages studied here are non-hysteretic and the transitions from Josephson current to the normal state are gradual. The observed smoothened IV curves can be understood by considering the noise

parameter that describes the ratio of the thermal energy to the Josephson coupling energy: $\Gamma = \frac{k_B T}{E_J} = \frac{2ek_B T}{I_0 \hbar}$. For a critical current of ~100 nA, this gives $\Gamma = $ ~1. It has been studied that in diffusive graphene Josephson weak links at millikelvin temperatures the supercurrent is reduced through a premature switching process[3,22,23] for $\Gamma << 1$. For stronger thermal fluctuation, however, phase diffusion dominates and causes a finite voltage in the supercurrent regime as well as gradual switching[24,25].

To our knowledge there is no analytical solution to the IV characteristics of a Josephson weak link under the strong thermal fluctuation with arbitrary $\beta_c$. To fit the IV curves and obtain the value of the intrinsic Josephson current $I_c$, we carry out a numerical simulation that solves the RCSJ model under a DC driving current in parallel with a Johnson–Nyquist noise current. To ensure the accuracy of our calculation, we compute the temperature dependence of the IV curves for the extreme case of a strongly overdamped junction where analytical solution was obtained using the Ambegaokar–Halperin model[25]. There our calculation precisely overlaps with the analytical solution (see Supplementary Information). We then apply the calculation to our device. In Fig. 3, the simulated IV curves are compared with the experimental data and quantitative agreements have been reached. In the fittings, the normal resistance outside the supercurrent regime is directly taken from the slope of the IV curves. The capacitance used here is fixed to 1.4 pF. For noise calculations we used the temperature $T = T_{EM} + T_b$, where $T_b$ is the actual bath temperature and $T_{EM}$ ~ 0.4 K is an estimated equivalent noise temperature of the electro-magnetic environment. The only free parameter used in the fittings is the Josephson current. From this we obtain the temperature and the gate voltage dependence of the Josephson current $I_c$, shown in Fig. 3 and Fig. 4. The Josephson current decreases with increasing temperature and with increasing normal resistance tuned by the gate voltage.

A very interesting feature of the intrinsic Josephson current is that except for near the Dirac point the Josephson current shows a linear dependence on the Fermi energy, as shown in Fig. 4b. This is in qualitative agreement with the theoretical prediction for a ballistic graphene Josephson weak link[2]. At large Fermi energies, such linear dependence can be related to the combination of a constant $I_cR_N$ product and a linear Fermi energy dependence of the normal conductance, which is a signature of the ballistic transport. Indeed, the observed the $I_cR_N$ product is roughly independent of the gate voltage except near the Dirac point. In Fig. 4a, significant decrease in the $I_cR_N$ product is evident for $V_g - V_D \leq 2V$ corresponding to low carrier density $n \leq 3 \times 10^{10} cm^{-2}$, well beyond the low-density limit for any on-substrate devices as a result of strong potential fluctuations. The observed decrease in the $I_cR_N$ product qualitatively agrees with the expectation of the quasi-diffusive transport of the evanescence modes near the Dirac point, predicted by the theory of the Josephson effect in ballistic graphene junctions[2]. On the other hand, at our base temperature of 2 K, the IV curves exhibit an Ohmic behavior for low carrier density n < $10^{10}$ cm$^{-2}$ as the Josephson energy decreases and the noise parameter becomes $\Gamma \sim 1$. More work is needed for the measurements at lower temperatures where lower supercurrent can be accurately measured at the close vicinity of the Dirac point. Compared with the theoretical predictions on ideal wide ballistic Josephson weak links[2,9] where $I_cR_N \sim 2.44\frac{\Delta}{e}$, the value of the $I_cR_N$ product obtained from our measurements, $\sim \frac{1}{3}\frac{\Delta}{e}$, is much smaller. We attribute the discrepancy to the imperfect graphene-superconductor interface and the narrow junction geometry (junction width ~ length).

Finally, to study the stability of the suspended graphene-NbN device, we thermal cycle our device by warming it to the room temperature and re-cooling it to repeat the measurements. The graphene-NbN device shows the excellent thermal stability and its quality is not affected by the thermal cycling. We also examine the device by placing it in raw vacuum (~100 mTorr) for approximately one month. The device has not shown signs of obvious aging after the idling time. Our results open a wide range of possibilities for future work, providing deeper understanding of the physics of the 2D Dirac fermion superconducting proximity effect as well as developing of novel superconducting devices.

**Acknowledgment**

The authors thank Dmitri Averin, Konstantin. K. Likharev, Daniel E. Prober, and Gleb Finkelstein for valuable discussions. We also thank Laszlo Mihaly for his support in cryogenic facilities. This work is supported by Air Force Office of Scientific Research Young Investigator Program, under Award FA9550-10-1-0090.

**Figure Captions**

**Figure 1. Fabrication of suspended graphene-NbN Josephson junction. a.** Graphene is mechanically exfoliated on top of PMMA resist and then spin-coated with MMA resist. **b.** E-beam lithography on the resists opens windows on the contact areas. **c.** NbN is deposited through reactive DC sputtering of Nb in Ar and $N_2$ plasma. **d.** Lift-off in solvents reveals the suspended graphene-NbN structure. **e.** False-color SEM image of a graphene-NbN Josephson junction. Here the junction width is W=1.8 μm, and length is L=0.5 μm.

**Figure 2. Effect of current annealing. a.** Carrier density dependence of mobility after current annealing. A maximum mobility of ~150,000 $cm^2$/Vs and a minimum carrier density of 5 x $10^9$ $cm^{-2}$ were measured on this device. To determine the mobility, the carrier density-gate voltage ratio, 1.72 x $10^{10}$ $cm^{-2}$/Volt, is calculated based on the geometry of the suspended device structure; here graphene is separated from the back gate by 285 nm of $SiO_2$ ($\varepsilon = 4$) and 250 nm of vacuum ($\varepsilon = 1$) (see Supplementary Information for more details). Inset: Comparison of resistance vs. $V_g$ before (black) and after (red) current annealing, measured at T > $T_c$. **b.** Evolution of the Multiple Andreev reflection characteristics before, during, and after current annealing.

**Figure 3. Supercurrent in suspended graphene-NbN Josephson junction. a.** Current-voltage characteristics at various gate voltages measured at T = 2.2 K. Symbols are experimental data and solid lines are corresponding fittings from numerical simulations. **b.** Temperature dependence of the supercurrent, measured at a gate voltage of $V_g$ = 5 V. Symbols are experimental data and solid lines are corresponding fittings from numerical simulations.

**Figure 4. Tuning of the intrinsic Josephson current. a.** Dependence of Josephson current (left axis) and $I_cR_N$ product on gate voltage $V_g$ - $V_D$ ($V_D$ being the gate voltage at the Dirac point) and carrier density. **b.** Dependence of Josephson current on Fermi energy. At large Fermi energies, the dependence is linear. Dotted line is given as a guide to the eyes for the linear dependence.

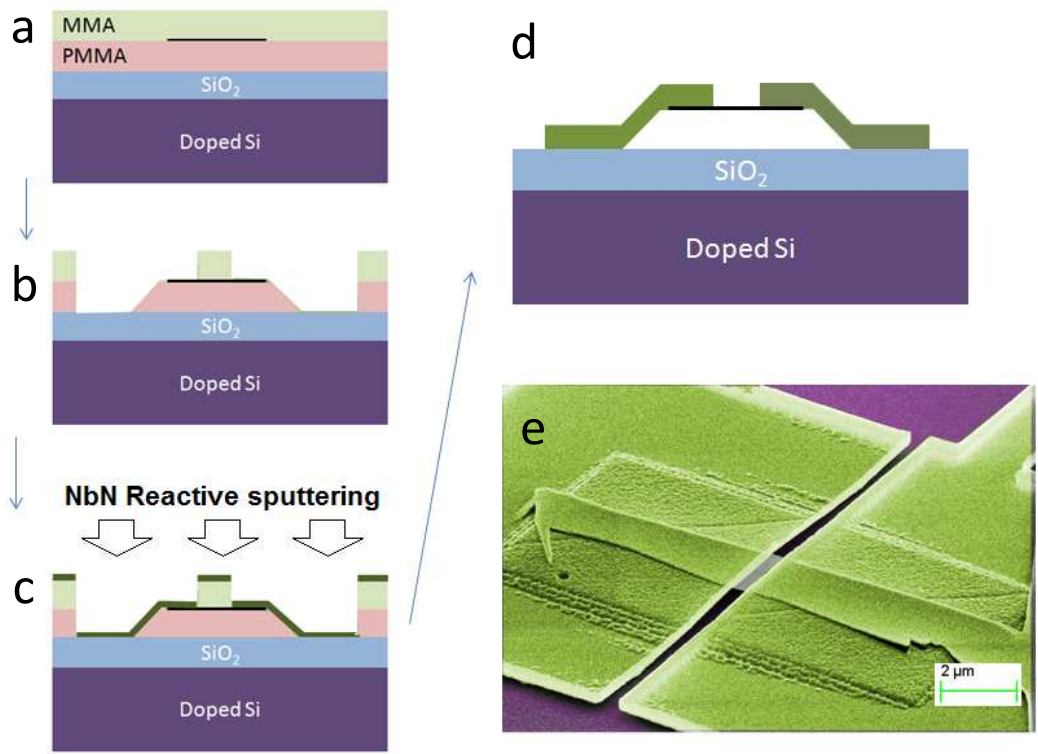

Figure 1.

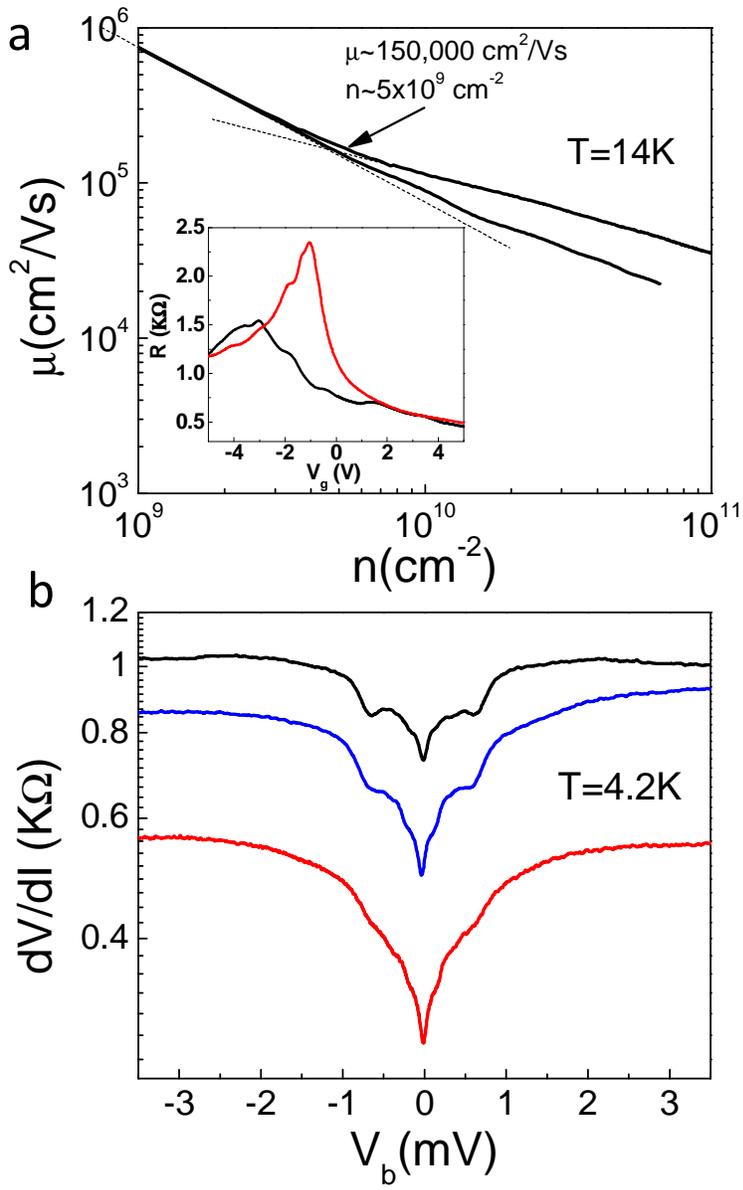

Figure 2.

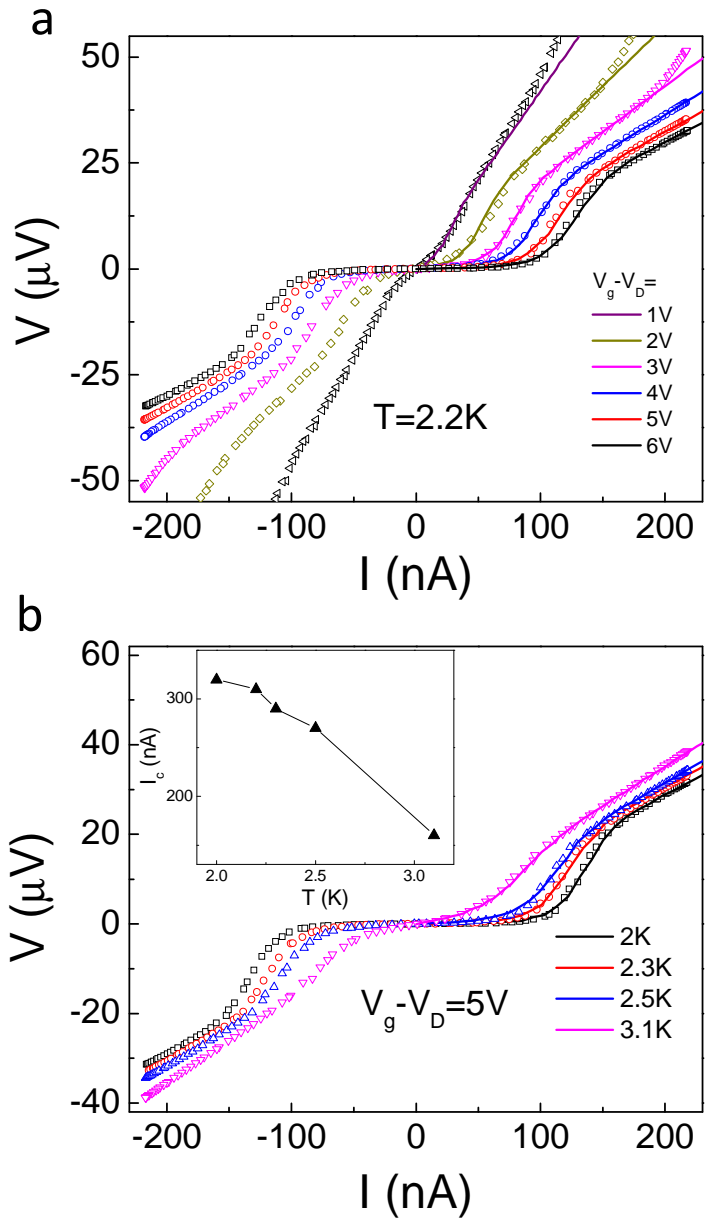

Figure 3.

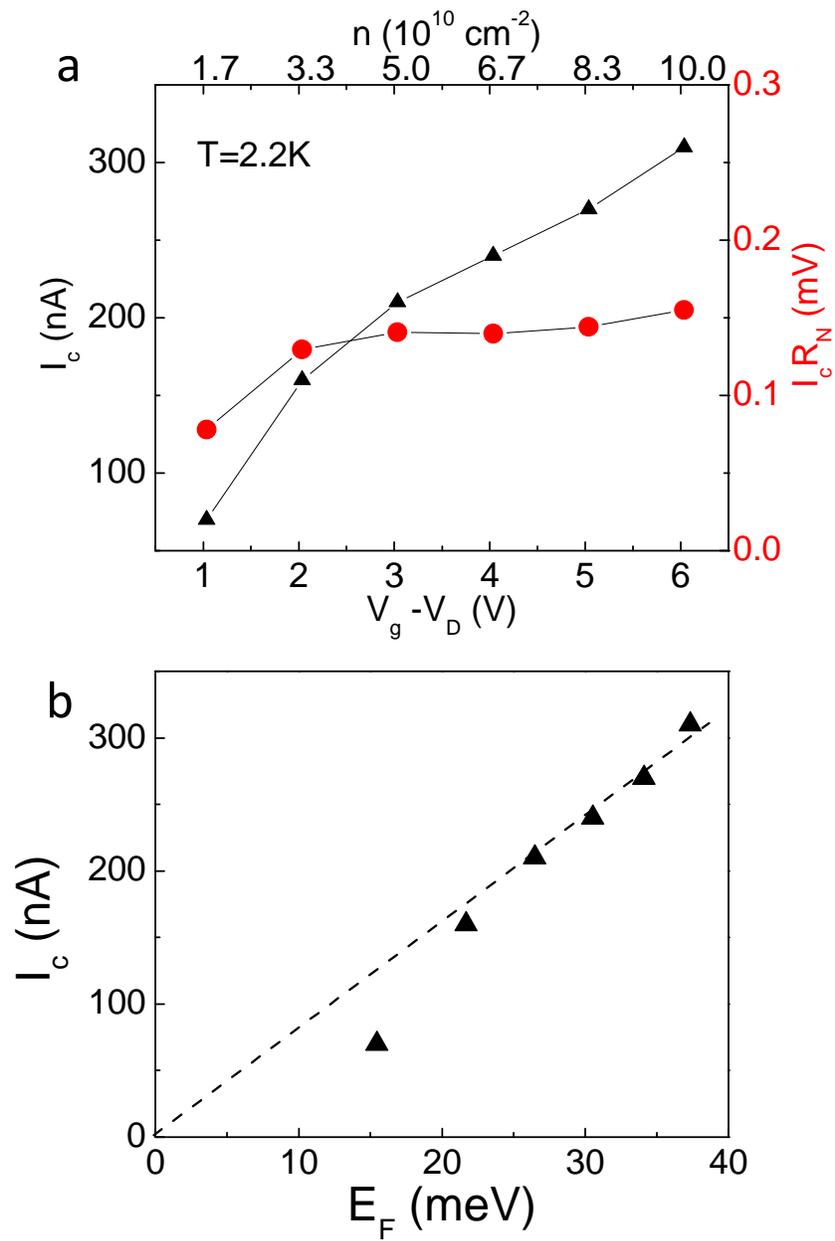

Figure 4.

**Suspended Graphene Ballistic Josephson Weak Links**

Naomi Mizuno, Bent Nielsen, and Xu Du

Department of Physics and Astronomy, Stony Brook University

**Supplementary Information**

*Contents*

- A. *Reactive sputtering of NbN for graphene Josephson devices*
- B. *E-beam lithography for suspended graphene devices*
- C. *Numerical Solution of the RCSJ model*

### A. Reactive sputtering of NbN for graphene Josephson devices

Two important factors affect the deposition of NbN films on graphene. The First is stress. We qualitatively determine the sign of the NbN film stress by sputtering the films on top of stress-free Al films pre-evaporated onto PMMA. In acetone, the lifted-off NbN/Al bi-metal films roll either upward or downward depending on the sign of the stress in NbN. Consistent with literature [26], it has been found that a low $Ar/N_2$ pressure results in a compressive stress, while a high $Ar/N_2$ pressure causes a tensile stress. We also notice that the NbN film can have the internal stress gradient, possibly caused by the change of local sputtering pressure/stoichiometry of the $Ar/N_2$ mixture. This effect is reduced by allowing a large flow of the gas mixture and at the same time reducing the impedance to achieve the same pressure. The reaction of $N_2$ with Nb causes the pressure to decrease during the sputtering. We use the pressure drop as a key parameter to determine if the flow is sufficient to produce uniformly stressed films. To achieve a preferred uniform stress, we find the pressure decrease to be less than 0.2 mTorr. A typical value of ~0.15 mTorr drop is used in our sputtering process.

The other important factor is the reaction of the adhesion Ti layer with the $N_2$ plasma. To understand the reaction between $N_2$ and the Ti sticking layer, we tested graphene devices at different metal deposition conditions. At a base pressure of ~2 x $10^{-8}$ Torr, Ti was evaporated onto graphene to form a 2 nm sticking layer. We then let in Ar/$N_2$ gases for 3 minutes, after which the gas mixture is pumped out. Finally, Au of 30 nm thickness is evaporated onto the device to form Au/Ti contacts. Following such metallization and lift-off procedure, the contact of the device has been measured to have a resistance of < 300 Ω*μm, similar to that of the typical Au/Ti contacts[27] without exposure to the Ar/$N_2$ gases after Ti evaporation. This suggests that Ar/$N_2$ itself does not affect the Ti sticking layer. By contrast, when we directly sputter NbN on the sample after the 2 nm Ti coating in Ar/$N_2$ plasma, the device shows much higher contact resistance of ~5 kΩ*μm, suggesting the immediate reaction of Ti in the Ar/$N_2$ plasma upon starting of the sputter gun. This result is further proved by the fact that the contact resistance is lowered by reducing the $N_2$ partial pressure during the reactive sputtering. To solve the problem of Ti-$N_2$ plasma reaction, we coat the Ti sticking layer with a 1.5 nm thick layer of Pd which is non-reactive to the $N_2$ plasma (here Pd is chosen for its formation of good contact with graphene by itself[27]). As a result we are able to recover 300 Ω*μm contact resistance. Another way to reduce the undesirable reaction between Ti and $N_2$ plasma is to reduce the $N_2$ concentration in the gas mixture. This can be done in two ways: One is to lower the initial $N_2$ partial pressure. The other is to baffle the pump which reduces the flow of Ar/N2 mixture for achieving the same sputtering pressure. While sputtering, the $N_2$ quickly reacts with the target that effectively lowers its concentration in the vicinity of the target/sample. Consequently the $N_2$ partial pressure drops significantly (0.5 - 0.9 mTorr). The decreased N ion density in the plasma reduces N-Ti reaction to the extent that the Ar ion bombardment reduction of TiN dominates over its formation. This way we can realize the improved contact resistance. The potential disadvantages of this technique include less controllability and reproducibility, also difficulty in obtaining reasonably high $T_c$ and low stress.

With considerations of the above factors, we choose the following condition for metallization; for sticking layer, we e-beam evaporate Ti (2 nm)/Pd (1.5 nm) at a base pressure of 2 x $10^{-8}$ Torr. Next, Ar (6.7 mTorr) and $N_2$ (0.9 mTorr) are introduced for sputtering at a fixed power of 470 W. The target is pre-sputtered for 80 s and then the sample is exposed to the plasma for NbN coating without turning off the sputter gun in between. 70 nm of NbN is coated onto the sample. During sputtering, the total pressure drops by ~0.15 mTorr as a result of Nb-$N_2$ reaction. The voltage and current values of the sputter gun are in the ranges of 314 - 319 V and 1.48 – 1.51 A, respectively.

### B. E-beam lithography for suspended graphene devices

E-beam lithography is performed on PMMA (bottom) / MMA (top) resist double layers, with graphene flakes sandwiched between the interfaces of the two layers. In our suspended graphene Josephson weak links, the graphene flakes are suspended by the contacts that form a 3-dimensional pill box-like structure. Such structure is made by controlling the e-beam exposure dose over the contact area. Near the graphene flake, we use a specific dose value so that only MMA layer will be completely developed. In contrast, we apply a higher dose on the outside support area to develop both MMA and PMMA. Between the two areas, we gradually change the exposure dose to form a slope-like transition region that gives a fine continuity of the contact metal thin films. For a thick PMMA under layer (hence graphene is suspended high above the surface of the substrate), such technique is needed to prevent formation of weak spots between the low dose and high dose areas. Fig. S1 shows a typical e-beam lithography design for the suspended graphene junctions with the gradual dose steps.

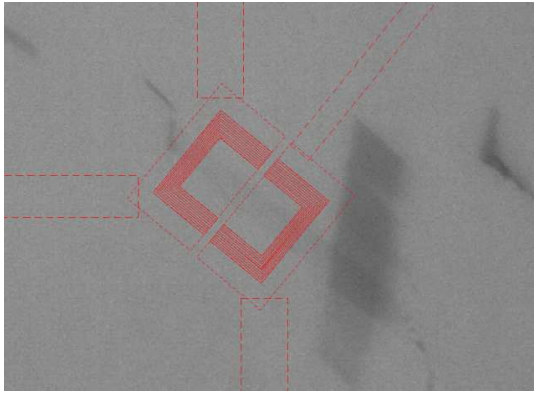

Figure S1. E-beam lithography design for the suspended graphene junctions.

After the e-beam exposure, the samples are developed in a MIBK : IPA (1 : 3) developer at 20 °C for 40 s, then immediately immersed in a water : IPA (1 : 3) developer at 5 °C for 60 s. The MIBK/ IPA mixture is used to develop the exposed MMA, while the water/IPA developer (which does not develop MMA) is used for the PMMA underneath. Over-developing of the top MMA layer is avoided by the use of the water : IPA developer, which significantly improves the smoothness of the slopes with the gradual dose change.

This chemical etchant-free e-beam resist double-layer technique can be applied to any substrate. The distance between graphene and the substrate surface can be estimated by the thickness of the PMMA resist with a good accuracy, as shown by the magnetotransport measurement in Fig. S2. Here, a graphene-Au/Ti junction, suspended by 250 nm-thick PMMA on a conducting substrate covered with 50 nm $Al_2O_3$, is measured in a 0.5 T magnetic field at T = 5.4 K. The high mobility of the device allows observation of quantum Hall oscillations corresponding to ν = 2, 6, 10, 14. From $\nu = \frac{nh}{eB}$ we obtain a carrier density to gate voltage ratio of 2.09 x $10^{10}$ $cm^{-2}$/V, corresponding to a graphene–surface distance of 257 nm that is reasonably close to the expected PMMA thickness.

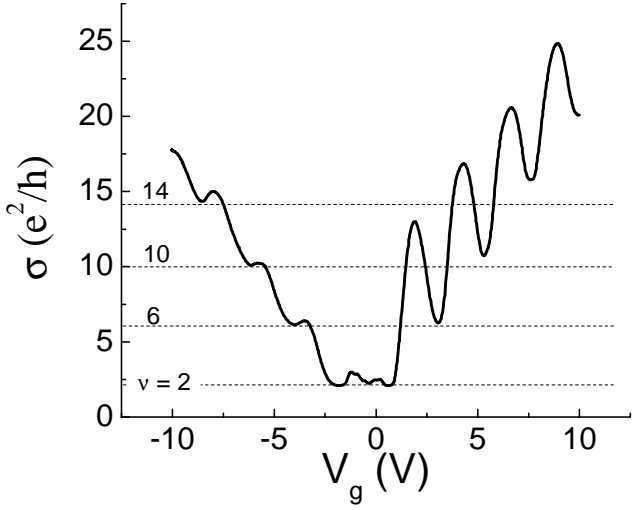

Figure S2. Magneto-oscillations of a suspended graphene–Au/Ti junction fabrication with the etchant-free method. The oscillation allows determination of the carrier density to gate voltage ratio.

### C. Numerical Solution of the RCSJ model

We assume that our devices can be described by the resistively and capacitively shunted Josephson junction (RCSJ) model, illustrated by the lumped-element circuit below.

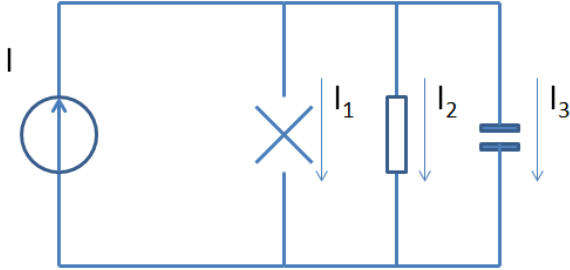

Figure S3. The RCSJ model.

The driving current (given) split into three flowing through the "pure" Josephson junction, the resistor and the capacitor. Hence;

$$i = \sin\phi + \frac{\hbar}{2eRI_c}\frac{d\phi}{dt} + \frac{\hbar C}{2eI_c}\frac{d^2\phi}{dt^2} \quad (1)$$

Here, $i = I/I_c = (I_{app} + I_{noise})/I_c$ consists of an applied current and a noise current.

To solve the equation numerically, we change the differential equation to difference equation:

$dt \to \Delta t$ and eq.(1) changes to:

$$\phi_{n+1} = \frac{1}{B}(i - \sin\phi_n)\Delta t^2 - \frac{A}{B}(\phi_n - \phi_{n-1})\Delta t + 2\phi_n - \phi_{n-1} \qquad (2)$$

Here $\Delta t$ is chosen to be much (~1000 times) smaller than the period of the Josephson oscillations. Given a initial condition of $\phi_1, \phi_2$, and the normalized current, $i(t)$, we can solve for $\phi(t)$. Then, from $\phi(t)$ we can calculate the averaged DC voltage: $\langle V \rangle = \frac{\hbar}{2e}\left\langle \frac{d\phi}{dt} \right\rangle$.

For $I_{noise}$ we used the Johnson noise current generated as a Gaussian white noise and related to the temperature by $\langle I_{noise}^2 \rangle = \frac{4k_B T}{R} f$, where $f$ is the bandwidth used in our simulation.

For each generated Johnson noise, we calculate an IV curve based on the above method. The IV curves are averaged 100 times over randomized Johnson noise currents. The final results are compared with the experimental data.

To ensure the reliability of the simulations, we compare our simulation results for the case of strongly underdamped junctions with the Ambegaokar – Halperin model [23,25], as shown in Fig. S4. The simulation results quantitatively agree with the model.

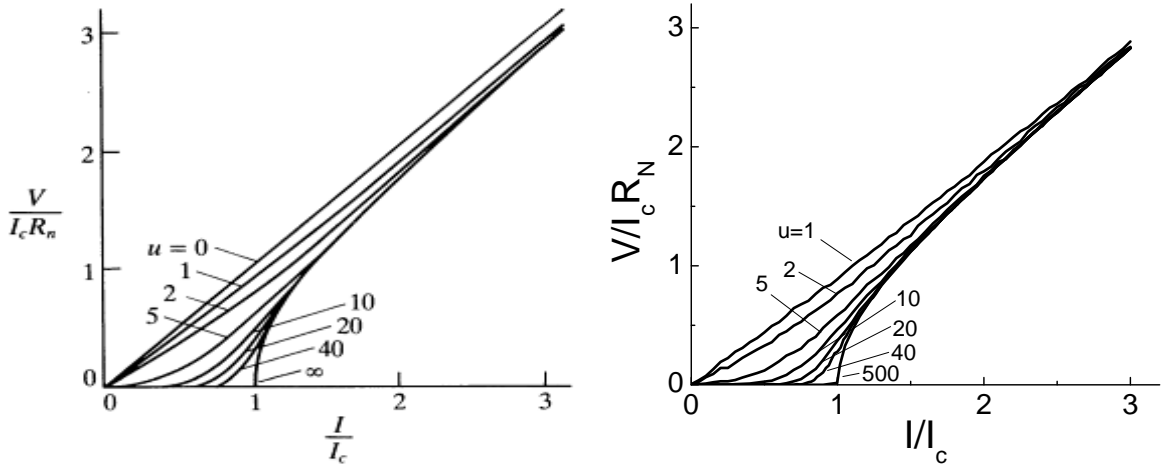

Figure S4. Side-by-side comparison of the Ambegaokar – Halperin theory for overdamped junctions (left) and our numerical simulations (right). Here $u = \hbar I_c / e k_B T$. In the simulation the Stewart-McCumber parameter $\beta_c = 0.05$. Here the left figure is taken from reference[23]. The two results precisely overlap with each other.